\documentclass[twocolumn,showpacs,preprintnumbers,amsmath,amssymb]{revtex4}
\usepackage{graphicx}
\usepackage[dvipsnames]{color}
\usepackage{bm}

\begin{document}

\title{UV active plasmons in alkali and alkaline earth intercalated graphene}

\author{V. Despoja$^{1,2}$}
\email{vito@phy.hr}
\author{L. Maru\v si\' c$^{3}$}
\email{lmarusic@unizd.hr}
\affiliation{$^1$Institut za fiziku, Bijeni\v{c}ka 46, 10000 Zagreb, Croatia}
\affiliation{$^2$Department of Physics, University of Zagreb, Bijeni\v{c}ka 32, HR-10000 Zagreb, Croatia}
\affiliation{$^3$Maritime Department, University of Zadar, M. Pavlinovi\'{c}a 1, HR-23000 Zadar, Croatia}

\begin{abstract}
The interband $\pi$ and $\pi+\sigma$ plasmons in pristine graphene and the Dirac plasmon in doped graphene are not applicable, since they are broad or weak, and weakly couple to an external longitudinal or electromagnetic probe. Therefore, the {\em ab initio} Density Function Theory is used to demonstrate that the chemical doping of the graphene by the alkali or alkaline earth atoms dramatically changes the poor graphene excitation spectrum in the ultra-violet frequency range ($4 - 10$ eV). Four prominent modes are detected. Two of them are the intra-layer plasmons with the square-root dispersion, characteristic for the two-dimensional modes. The remaining two are the inter-layer plasmons, very strong in the long-wavelength limit but damped for larger wave-vectors. The optical absorption calculations show that the inter-layer plasmons are both optically active, which makes these materials suitable for small organic molecule sensing. This is particularly intriguing because the optically active two-dimensional plasmons have not been detected in other materials.
\end{abstract}

\maketitle

Extensive research of electronic excitations in graphene showed the existence of several two-dimensional (2D) plasmon modes: the intraband (Dirac) plasmon existing only in the doped graphene \cite{DasSarma,PlPhT,grafen,Measurenanoribb,Tip,IR2}, and the interband plasmons, which exist in pristine and doped graphene and originate from the interband electron-hole transition between the $\pi$ and $\pi^*$ bands and between the $\pi$ and $\sigma^*$ bands \cite{grafen,politano1,politano2,Dino,Eberlein}. These investigations also showed that the interband $\pi$ and $\pi+\sigma$ plasmons are broad and weak resonances, so their interaction with the external longitudinal or electromagnetic probes is weak as well, which makes them inadequate for most practical applications. The 'tunable' Dirac plasmon in the doped graphene is also weak (for experimentally feasible doping), and in addition to that, it does not couple to an incident electromagnetic field directly. In the systems proposed so far, light could be coupled to the Dirac plasmon only indirectly, e.g. by using metallic tips, gratings or prisms, or by arranging graphene into nanoribbons\cite{Measurenanoribb,Tip,PRLGNR}, which is all hard to fabricate. Also, such indirect coupling additionally reduces the intensity of the plasmon, thus reducing the efficiency of its application. 

The alkali or alkaline earth intercalated graphene is much easier to fabricate and offers a broader variety of plasmons, both intraband and (especially) interband. Such systems have recently been extensively studied, both theoretically and experimentally \cite{exp1,exp2,exp3,exp1SuperC,exp2SuperC,theory,lazic}, but the attention has not been on the electronic excitations. Intercalating any alkali or alkaline earth metal to a single graphene layer causes the natural doping of the graphene and results in the formation of two quasi two-dimensional (q2D) plasmas. This supports the existence of two 2D intraband plasmons, acoustic and Dirac, with frequencies up to 4 eV \cite{2Dplasmons}, as well as several interband and even inter-layer modes occurring at higher frequencies. Some of these modes are optically active and some of them can be manipulated by doping, which opens possibilities for their application in various fields, such as plasmonics, photonics, transformation optics, optoelectronics, light emitters, detectors and photovoltaic devices \cite{IR3,APP1,chinos,appl1,appl3,appl4,appl5,appl6,appl7,appl8}. Moreover, 'tunable' 2D plasmons could be very useful in the area of chemical or biological sensing \cite{APP2,photopto,appl2,appl9}, which is one of our main suggestions for the potential application of the results of this research.  

We performed calculations for several alkali and alkaline earth metals, with different coverages, and found that the effects which are the focus of this letter are valid for all of them. In all these cases, in addition to the graphene $\pi$ and $\sigma$ bands, there are also the $\pi$ and $\sigma$ bands of the intercalated metal. This opens possibilities for various electron-hole (e-h) transitions which may be the origins of the interband plasmons. We limit our investigation to the frequencies between 4 and 10 eV(the UV region), where the dominant interband plasmons occur, and identify four significant modes within this range. Two of them are not very well defined in the long-wavelength limit but they exist at larger wave-vectors as well, and show the square-root dispersion characteristic for the surface and 2D modes. These modes are the intra-layer modes, one located in the graphene layer and the other located in the metallic layer. The other two are very prominent in the long-wavelength limit, but at higher wave-vectors their intensities rapidly decrease, which makes them potentially interesting for optical applications \cite{IR3,APP1,APP2,stauber,PRLGNR}. Their dispersions are different from those typical for the 2D modes, indicating that they are different from the usual 2D plasmons. Detailed inspection (including retardation, i.e. finite speed of light, and tensorical response) shows that they are dipolar inter-layer modes (the electric field they produce oscillates perpendicular to the crystal plane), i.e. optically active q2D plasmons, contrary to the widely studied q2D plasmons which produce electric field parallel to the crystal plane, and are not optically active. The extensively studied graphene $\pi$ and $\pi+\sigma$ modes are optically active, but in the long wavelength limit ($Q \rightarrow 0$) they are not plasmons but electron-hole excitations \cite{Dino}.

The theoretical formulation of the electronic response in various q2D systems has already been presented\cite{grafen,Duncan2,wake,Rukelj}, so here we only point out some details of the calculation important for the understanding of the result we want to present. We define the Electron Energy Loss Spectroscopy (EELS) local spectral function as the imaginary part of the excitation propagator
\begin{equation}
S_{z_0}({\bf Q},\omega)=-Im D_{z_0}({\bf Q},\omega),
\label{spectrum}
\end{equation}
where
\begin{equation}
D_{z_0}({\bf Q},\omega) = W^{ind}_{\textbf{G}_{\parallel}=0}(\textbf{Q},\omega,z_0,z_0).
\label{propagator}
\end{equation}

The $S_{z_0}({\bf Q},\omega)$ is also proportional to the probability density for the parallel momentum transfer ${\bf Q}=(Q_x,Q_y)$ and the energy loss $\omega$ of the reflected electron in the Reflection Electron Energy Loss Spectroscopy (REELS)\cite{REELS}. The induced dynamically screened Coulomb interaction is $W^{ind}=v^{2\textrm{D}}\otimes\chi\otimes v^{2\textrm{D}}$, where $v^{2\textrm{D}} = \frac{2\pi}{Q}e^{-Q\left|z-z'\right|}$ is the 2D Fourier transform of the bare Coulomb interaction and $\otimes=\int^{L/2}_{-L/2}dz$\cite{Leo}. The response function is obtained as the solution of the matrix Dyson equation $\hat{\chi}=\hat{\chi}^0 +\hat{\chi}^0\hat{v}^{2\textrm{D}}\hat{\chi}$ in the reciprocal space plane-wave basis ${\bf G}=({\bf G}_{\parallel},G_z)$. The non-interacting electrons response matrix is $\hat{\chi}^{0}=\frac{2}{\Omega}\sum_{i,j}(f_i-f_j)/(\omega+i\eta+E_i-E_j)\rho_{{\bf{G}},ij}\rho^*_{{\bf G}',ij}$, where $f_i$ is the Fermi-Dirac distribution, $\rho_{{\bf{G}},ij}$ are charge vertices ~\cite{grafen}, $\Omega$ is the normalization volume, and $i=(n,\bf{K})$ and $j=(m,{\bf K+Q})$ are Kohn-Sham-Bloch states. The Coulomb interaction with the surrounding supercells in the superlattice arrangement is excluded, as described in detail in Ref.\cite{Rukelj}.

\begin{widetext}

\begin{figure}[h]
\centering
\includegraphics[width=\textwidth]{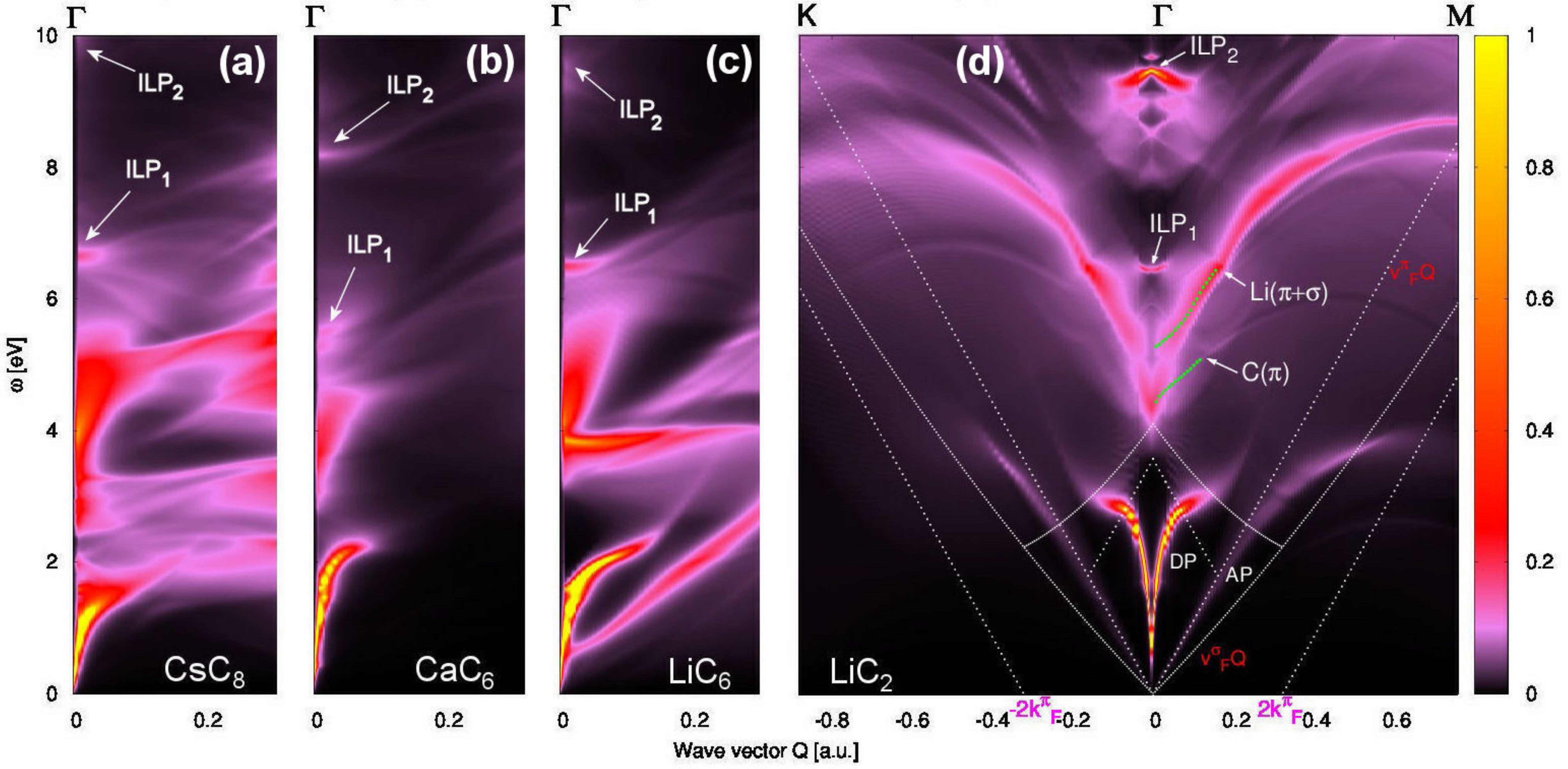}
\caption{(color online) The intensity of the electronic excitations in (a) CsC$_8$, (b) CaC$_6$, (c) LiC$_6$ and (d) LiC$_2$. The white and green dotted lines in (d) show the boundaries of the e-h excitation gaps for the graphene $\pi$ bands around the Dirac point and the Li $\sigma$ bands around the $\Gamma$ point, respectively.}
\label{Fig1}
\end{figure}

\end{widetext}

To calculate the Kohn-Sham (KS) wave functions $\phi_{n{\bf K}}$ and energy levels $E_{n{\bf K}}$, i.e. the band structure, of the LiC$_2$, LiC$_6$, CaC$_6$ and CsC$_8$ slabs, we use the plane-wave self-consistent field DFT code (PWSCF) within the QUANTUM ESPRESSO (QE) package \cite{QE}. The core-electron interaction is approximated by the norm-conserving pseudopotentials \cite{pseudopotentials}, and the exchange correlation (XC) potential by the Perdew-Zunger local density approximation (LDA) \cite{LDA}. For the slab unit cell constant we use the graphene value of $a_{uc} = 4.651$ a.u. \cite{lattice}, and we separate the slabs by $L=5a_{uc}=23.255$a.u. The equilibrium separations between the metallic and carbon layers within a slab for these four systems is $d = $ 4.1a.u.(2.17\AA), 3.28a.u.(1.74\AA), 4.46a.u. (2.36\AA) and 5.8a.u. (3.08\AA), respectively, as proposed in Ref.\cite{lazic,Nanolett-Dino}. Our reference frame is chosen so that the graphene layer is positioned at $z=0$, and the metallic layer is at $z=d$. The ground state electronic densities of the slabs are calculated by using the $12\times12\times1$ Monkhorst-Pack K-point mesh\cite{MPmesh} of the first Brillouin zone (BZ). For the plane-wave cut-off energy we choose $50$Ry ($680$eV). The Fermi levels of these systems (measured from the Dirac point, i.e. from the pristine graphene Fermi level) are: $E_F$ = $1.78$, $1.55$, $1.375$ and $1.24$eV, respectively. For the response matrix $\hat{\chi}^{0}$ calculation in the long-wavelength ($Q<0.01$a.u.) limit we use $601\times601\times1$ $K$-point mesh and the damping parameter $\eta=10meV$, while for the larger $Q$s we use $201\times201\times1$ $K$-point mesh sampling and $\eta=30meV$. In all cases the band summation is performed over $30$ bands and the perpendicular crystal local field energy cut off is $10$Ry ($136$eV), which corresponds with $23$ $G_z$ wave vectors. 

Figs. \ref{Fig1} shows the excitation spectra in (a) CsC$_8$, (b) CaC$_6$, (c) LiC$_6$ and (d) LiC$_2$ slabs, calculated from Eg.\ref{spectrum}. The spectral intensities are shown as functions of $\omega$ and Q, using the color scheme, which enables us to see the dispersions of the modes. We can see that, in addition to the well known modes present in the doped graphene (Dirac (DP) and $\pi$ (C($\pi$)) plasmon\cite{grafen}), there are a few other modes, strong in the long-wavelength limit (indicating their optical activity) and more pronounced in the systems with higher electronic doping, especially for LiC$_6$ and LiC$_2$. Therefore, we put emphasis on the system with the highest doping, i.e. to the LiC$_2$. Fig.\ref{Fig1}(d) shows the intensities of the electronic excitations in the $\Gamma$M and $\Gamma$K direction for the LiC$_2$. The spectra in these two directions are very similar, so here we focus only on the $\Gamma$M direction. At lower frequencies (up to 4 eV) we can see the intra-band q2D plasmons, which have already been discussed in detail for the LiC$_2$\cite{2Dplasmons}. At frequencies between 4 and 10 eV we can see four significant inter-band modes (denoted as C($\pi$), Li($\pi + \sigma$), ILP$_1$ and ILP$_2$), two with the square-root dispersion, characteristic for the q2D systems, which exist for the larger wavevectors as well, and the other two which are strongly damped for larger wavevectors. These modes, which exist in all these systems (at similar frequencies), are the focus of this letter. To understand them we shall explore the band structure and the spectra of electronic excitations in the LiC$_2$ in more detail. However, our conclusions about the origins and characteristic of the modes obtained for the LiC$_2$, are valid for the other three systems as well.

\begin{figure}[h]
\centering
\includegraphics[width=0.45\textwidth]{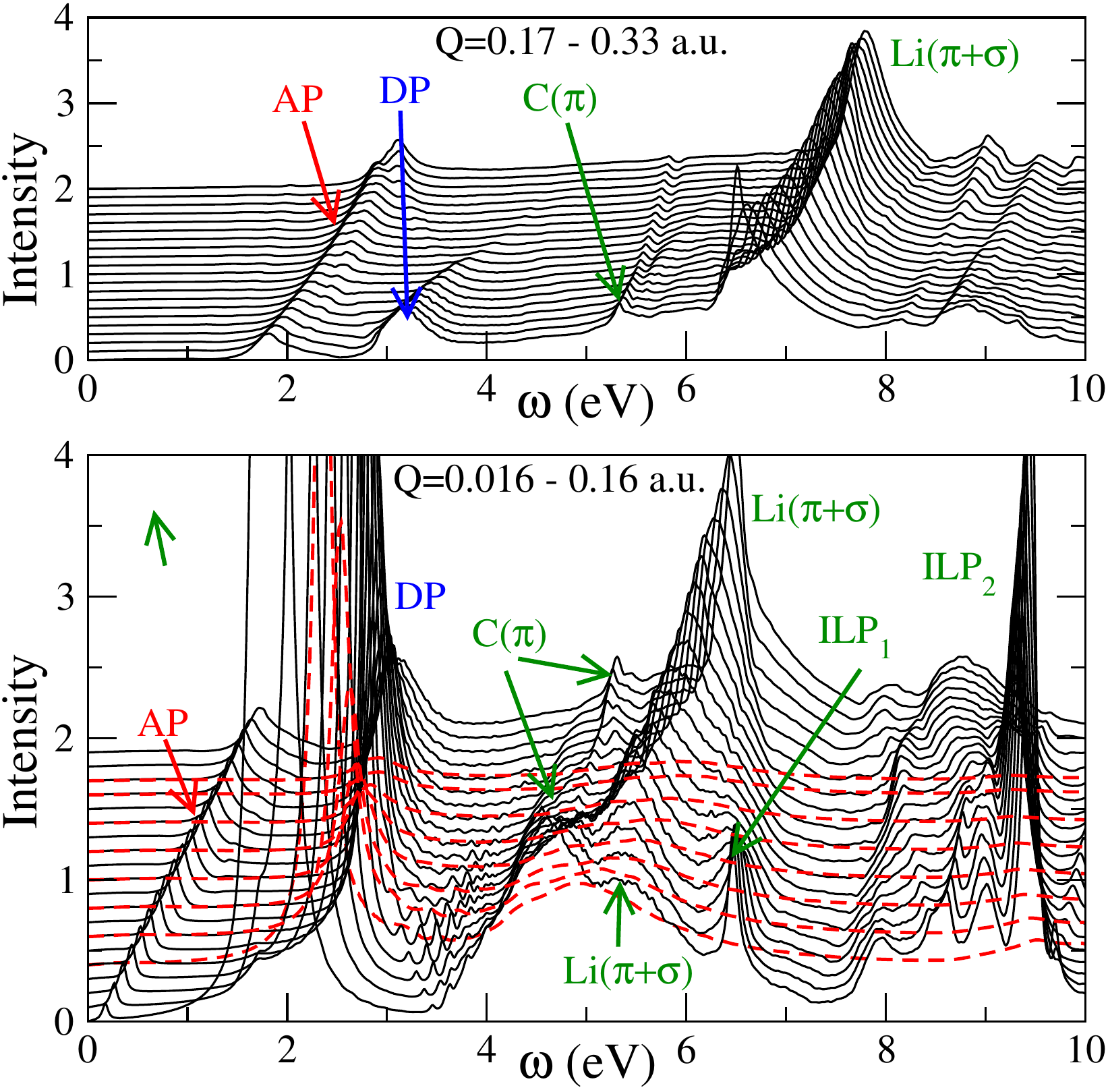}
\caption{(color online) The spectra of the electronic excitations in the LiC$_2$ for the $\Gamma$M direction (full black lines). The lower panel contains the comparison with the doped graphene spectra (dashed red lines) with the matching Fermi levels. Wave-vector ranges for each figure are indicated on the figure.}
\label{Fig2}
\end{figure}

Fig.\ref{Fig2} shows spectra $S(\omega)$ for the LiC$_2$ for various wave-vectors Q (denoted in graphs) in the $\Gamma$M direction. The lower panel contains the spectra of the n-doped graphene (dashed red lines), with the same doping as in the LiC$_2$ ($E_F=1.78$eV), for comparison. The doped graphene spectra show only two modes, the very prominent intraband Dirac plasmon, roughly matching the LiC$_2$ Dirac plasmon, and the interband $\pi$ plasmon around 5eV, which is very weak due to heavy doping. In the LiC$_2$ spectra, in addition to the already described q2D intraband acoustic and Dirac plasmon (AP and DP)\cite{2Dplasmons}, we can notice a barely visible broad peak between 4.5 and 5eV, which corresponds with the graphene $\pi$ plasmon, plus three other modes which cannot be related to any of the graphene modes. This means that these modes are either the lithium q2D intra-layer modes, or the inter-layer modes, which represent charge oscillations perpendicular to the crystal plane.

\begin{widetext}

\begin{figure}[h]
\centering
\includegraphics[width=0.29\textwidth]{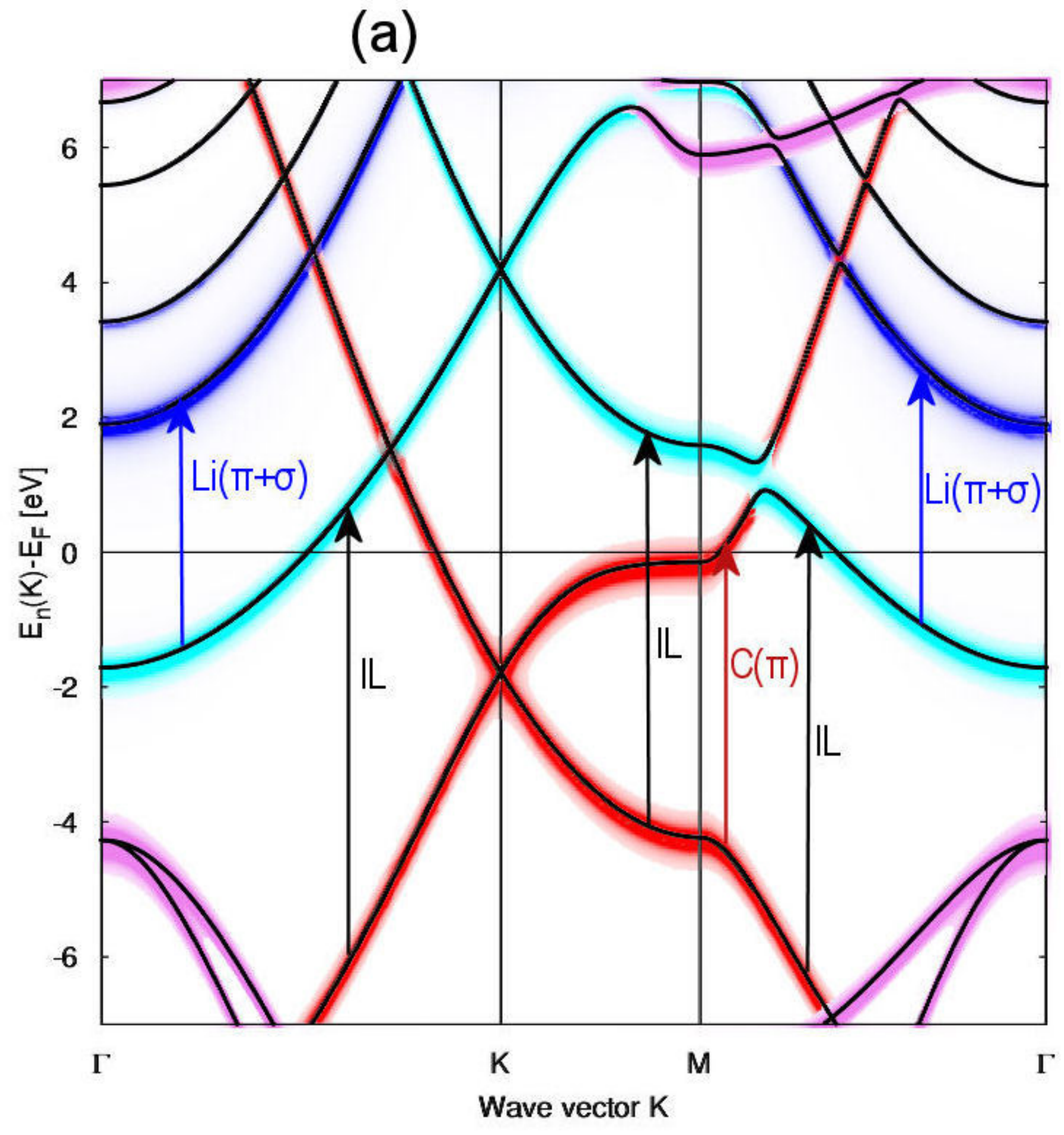}
\hspace{5mm}
\includegraphics[width=0.4\textwidth]{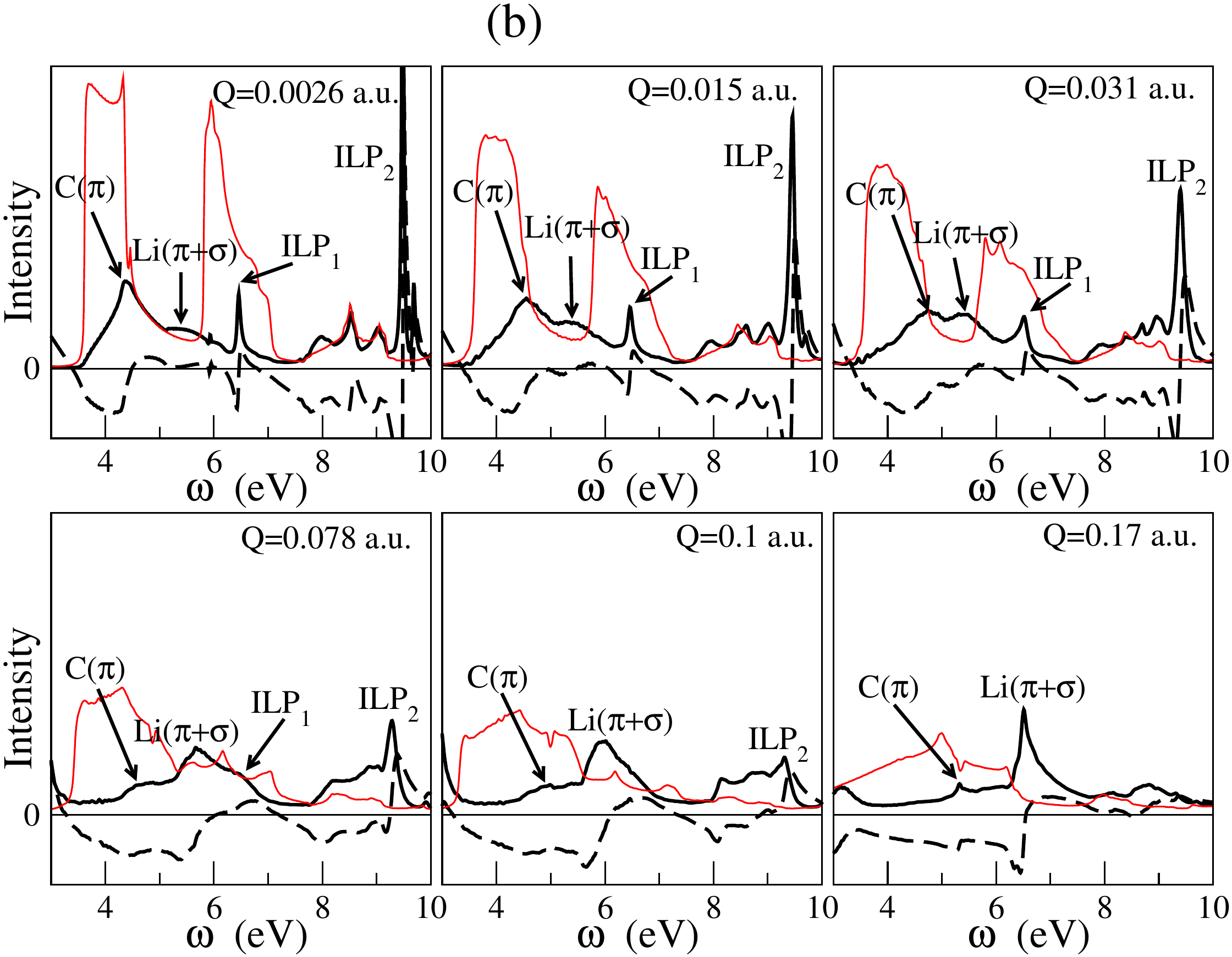}
\hspace{5mm}
\includegraphics[width=0.22\textwidth]{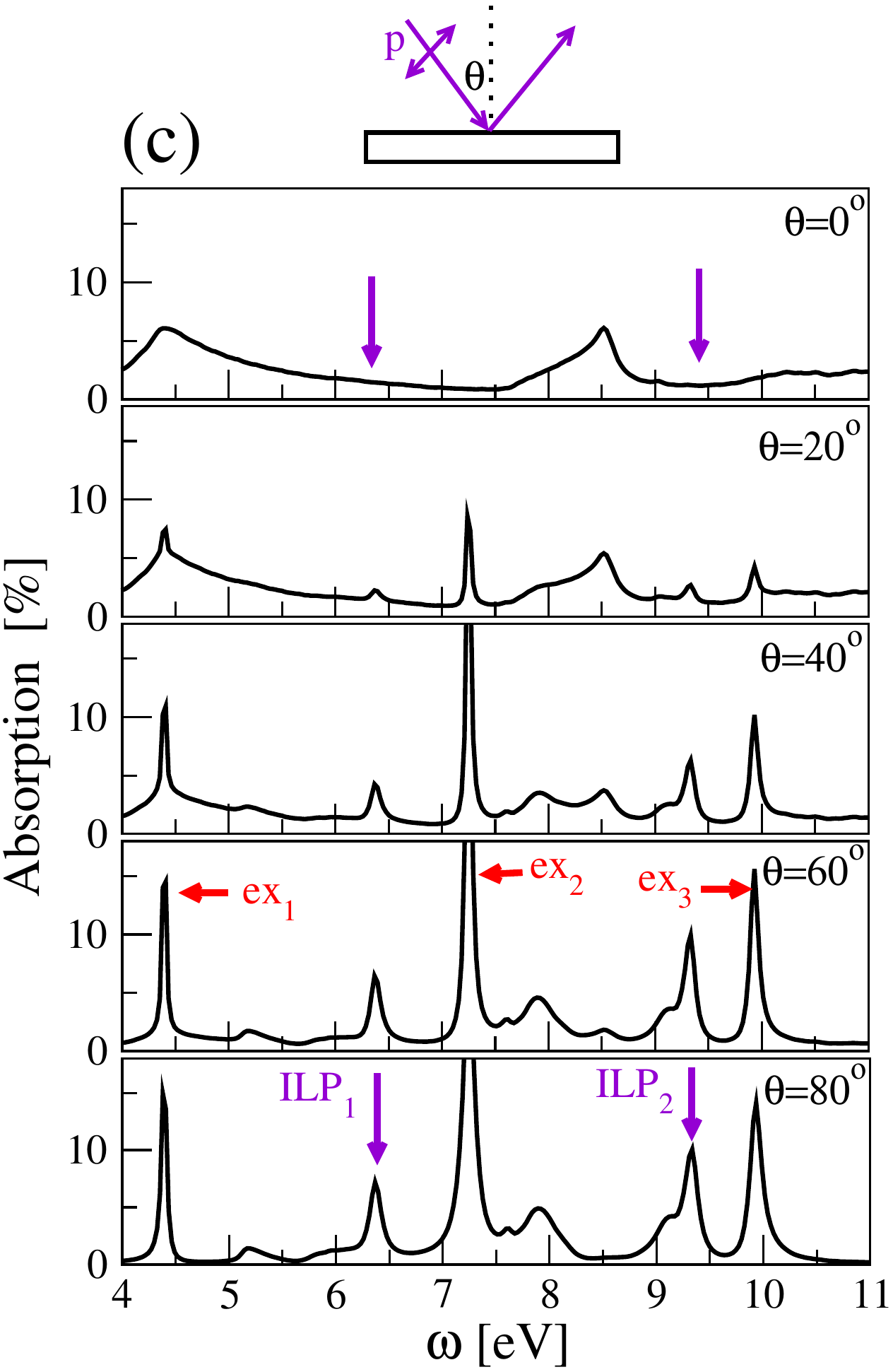}
\caption{(color online) (a) The band structure of the LiC$_2$, with the color scheme indicating the predominant origins of particular bands (blue - Li($\pi$), turquoise - Li($\sigma$), pink - C($\sigma$), red C($\pi$)).(b)  Development of the interband plasmons with the increase of the wavevector. Thick full black lines - $Im D$, thick dashed black lines - $Re D$, thin red lines - spectra of the single particle excitations only. (c) Angle resolved optical absorption spectra in LiC$_2$, in ultraviolet frequency range.}
\label{Fig3}
\end{figure}

\end{widetext}

Fig.\ref{Fig3}(a) shows the band structure of the LiC$_2$ slab, with the color scheme indicating the predominant origins of particular bands. Blue and turquoise indicate predominant lithium $\pi$ and $\sigma$ orbitals, respectively, while red and pink indicate predominant graphene $\pi$ and $\sigma$ orbitals, respectively. The arrows indicate the e-h transitions which are the potential origins of the four modes. However, since the transition energies are very similar, it is impossible to reach definite conclusions about the origins of the particular modes from the band structure itself. Fig.\ref{Fig3}(b) shows the imaginary (thick solid black line) and real (thick dashed black line) part of the excitation propagator (\ref{propagator}) in the LiC$_2$ for several characteristic wave-vectors $Q$. The thin red line is the unscreened (single particle) spectrum obtained by replacing $\chi$ with $\chi^0$ in $W^{ind}$ used in (\ref{spectrum}). By comparing these three lines we can distinguish the collective modes from the single particle excitations. Furthermore, comparing the frequencies of these excitations with the transitions in the band structure in the Fig.\ref{Fig3}(a) can help us identify the origins of some of the modes.

In the unscreened spectra (thin red lines) we can see two prominent peaks, one around 4ev which exist for all wavevectors, and the other around 6eV which disappears for larger Q. This indicates that the first one is the origin of the modes denoted as C($\pi$) and Li($\pi+\sigma$), while the second one is the origin of the remaining two modes (IL$_1$ and IL$_2$). Further analysis consists of applying the p and n doping to our system (i.e. changing the position of the Fermi level and causing some occupied bands to become unoccupied, and vice versa), and omitting particular bands from the calculation of the response function ${\chi}^{0}$, to determine the exact role of each band. By doing that, we found out that the first peak in the single particle spectra is actually an overlap of two peaks. One is coming from the transition between the $\pi$ and $\pi^*$ graphene bands around the M point (red arrow in Fig.\ref{Fig3}(a)), and that one is the origin of the mode denoted as C($\pi$), i.e. the graphene $\pi$ plasmon. The other is coming from the transition between the $\sigma$ and $\pi$ lithium bands around the $\Gamma$ point (blue arrow in Fig.\ref{Fig3}(a)), and that one is the origin of the mode denoted as Li($\pi+\sigma$), i.e. the lithium $\pi+\sigma$ plasmon. The second peak in the single particle spectra comes from the transitions between the graphene $\pi$ bands and the lithium $\sigma$ bands (black arrows in Fig.\ref{Fig3}(a)), and it is the origin of the remaining two modes, denoted as IL$_1$ and IL$_2$. Their dispersion is not square root like, which is the consequence of their inter-layer nature. All the presented spectra are calculated for the probe positioned at $z_0=L/2$, but we can change the position of the probe and monitor the changes in the spectra to determine the symmetry of the particular modes. By doing that we confirmed that one of the peaks (IL$_1$) is even, while the other one (IL$_2$) is odd.

Contrary to the intra-layer plasmons C($\pi$) and Li($\pi+\sigma$) the inter-layer plasmons IL$_1$ and IL$_2$ are sharp, well defined resonances which could be especially suitable for the sensing of small organic molecules with excitonic spectra in UV frequency range. However, the crucial question is: can the IL$_1$ and IL$_2$ plasmons be excited by an external electromagnetic field, i.e, are they optically active? If that is the case, then it seems that the intercalated graphene may become the technologically simplest platform for biosensing. In the systems proposed so far, light could be coupled to plasmon resonances only indirectly, e.g. by using the metallic nanoparticles, gratings or prisms, or by arranging graphene into nanoribbons, which is much more difficult to fabricate. Another exciting aspect of this issue is that the optically active q2D plasmons have not been discovered in other systems. In order to answer this crucial questions we performed a sophisticated angle resolved optical absorption calculation which includes the retardation and the tensoric character of the LiC$_2$ dynamical response \cite{DSV}. The mathematical formulation of this theoretical tool and its application to the molybdenum disulfide (MoS$_2$) monolayer is presented in Ref.\cite{MoS2}.

Fig.\ref{Fig3}(c) shows the absorption of $p$ polarized light in the LiC$_2$, as a function of the incident light with the frequency $\omega$ and angle $\theta$ (as sketched). For the normal incidence ($\theta=0^o$) the electric field is parallel with the crystal plane and there are no peaks corresponding to IL$_1$ and IL$_2$. However, as the incident angle increases the IL$_1$ and IL$_2$ peaks appear, and finally for the almost grazing incidence ($\theta=80^o$), i.e. for the the electrical field almost perpendicular to the crystal plane, they become very intensive. This undoubtedly confirms not only that these modes are optically active, but also their inter-layer character. The grazing spectra show some additional peaks (ex$_1$, ex$_2$ and ex$_3$) which do not appear in the EELS spectra, i.e. which can not be excited by an external longitudinal probe, which means that they are probably not plasmons but UV active excitons.   
 
In the remaining three systems presented in this letter, the electronic doping is weaker than in the LiC$_2$ which has two important consequences. First, the two inter-layer plasmons are not nearly as strong and sharp as they are in the LiC$_2$, but they still exist, and are still optically active in the UV region. Second, due to the weaker doping the Fermi level is lower (with respect to the LiC$_2$ Fermi level) by 0.23eV for LiC$_6$, 0.405eV for CaC$_6$ and 0.54eV for CsC$_8$. Considering that the plateau in the graphene $\pi^*$ band is only 0.1 - 0.2eV below the LiC$_2$ Fermi level, this means that in the three other systems that plateau is unoccupied, which makes the graphene $\pi$ plasmon much stronger. Therefore, by changing the doping (by changing the dopant or the coverage, or by applying the gate voltage) we can tune these modes, i.e. increase or decrease their intensities.

In conclusion, we showed that doping the graphene by the alkali or alkaline earth atoms dramatically modifies the graphene plasmonics, especially in UV parts of the spectra, where we obtain four interband plasmons. This effect is the strongest in the full coverage lithium doped graphene (LiC$_2$), due to the highest doping. Two of the modes, not very strong in the long wavelength limit, exist for larger wavevectors, with the square-root dispersion characteristic for the 2D plasmons. They turned out to be the intra-layer modes, one within the graphene layer (the well known graphene $\pi$ plasmon), and the other within the intercalated metal layer. The other two plasmons  ILP$_1$ and ILP$_2$ are strong and sharp in the long wavelength limit, but damped for the larger wave-vectors. They turned out to be the inter-layer optically active plasmons, i.e. they couple directly to the electromagnetic field. Such unusual and poorly explored optically active 2D plasmons can be used as the efficient sensor in chemical sensing and biosensing. 

\begin{acknowledgements}
This work was supported by the QuantiXLie Centre of Excellence, a project co-financed by the Croatian Government and European Union through the European Regional Development Fund - the Competitiveness and Cohesion Operational Programme (Grant KK.01.1.1.01.0004). Computational resources were
provided by the Donostia International Physic Center
(DIPC) computing center.
\end{acknowledgements}


\begin{thebibliography}{99}
\bibitem{DasSarma} E. H. Hwang and S. Das Sarma, Phys. Rev. B {\bf 75}, 205418 (2007), Hwang E. H. and Das Sarma, Phys. Rev. B  {\bf 80}, 205405 (2009)
\bibitem{PlPhT} E. H. Hwang, R. Sensarma, and S. Das Sarma, Phys. Rev. B {\bf 82}, 195406 (2010)
\bibitem{grafen} V. Despoja, D. Novko, K. Dekani\'{c}, M. \v{S}unji\'{c} and L. Maru\v{s}i\'{c}, Phys. Rev. B {\bf 87}, 075447 (2013)
\bibitem{Measurenanoribb} H. Yan, T. Low, W. Zhu, Y. Wu, M. Freitag, X. Li, F. Guinea, P. Avouris and F. Xia, Nat Photonics {\bf 7}, 394 (2013)
\bibitem{Tip} Z. Fei, G. O. Andreev, W. Bao, L. M. Zhang, A. S. McLeod, C. Wang, M. K. Stewart, Z. Zhao, G. Dominguez, M. Thiemens, M. M. Fogler, M. J. Tauber, A. H. Castro-Neto, C. N. Lau, F. Keilmann and D. N. Basov, Nano Lett. {\bf 11}, 4701 (2011)
\bibitem{IR2} M. Jablan, H. Buljan and M. Solja\v ci\' c, Phys. Rev. B {\bf 80}, 245435 (2009)
\bibitem{politano1} A. Politano, I. Radovi\'{c}, D. Borka, Z.L. Mi\v{s}kovi\'{c}, G. Chiarello, Carbon {\bf 96}, 91 (2016)
\bibitem{politano2} A. Politano, I. Radovi\'{c}, D. Borka, Z.L. Mi\v{s}kovi\'{c}, H.K. Yu, D. Far\'{i}as, G. Chiarello, Carbon {\bf 114}, 70 (2017)
\bibitem{Dino} D. Novko, V. Despoja, and M. \v{S}unji\'{c}, Phys. Rev. B {\bf 91}, 195407 (2015)
\bibitem{Eberlein} T. Eberlein, U. Bangert, R. R. Nair, R. Jones, M. Gass, A. L. Bleloch, K. S. Novoselov, A. Geim, and P. R. Briddon, Phys. Rev. B {\bf 77}, 233406 (2008)
\bibitem{PRLGNR} C. Vacacela Gomez, M. Pisarra, M. Gravina, J.M. Pitarke, and A. Sindona, Phys. Rev. Lett. {\bf 117}, 116801 (2016)
\bibitem{exp1} A. Kumar, A. L. M. Reddy, A. Mukherjee, M. Dubey, X. Zhan, N. Singh, L. Ci, W. E. Billups, J. Nagurny, G. Mital and P. M. Ajayan, ACS Nano {\bf 5}, 4345 (2011) 
\bibitem{exp2} S. L. Yang, J. A. Sobota, C. A. Howard, C. J. Pickard, M. Hashimoto, D. H. Lu, S. K. Mo, P. S. Kirchmann and Z. X. Shen, Nature Communications {\bf 5}, 3493 (2014)
\bibitem{exp3} N. M. Caffrey, L. I. Johansson, C. Xia, R. Armiento, I. A. Abrikosov and C. Jacobi, Phys. Rev. B {\bf 93} 195421 (2016)
\bibitem{exp1SuperC} S. Ichinokura, K. Sugawara, A. Takayama, T. Takahashi, and S. Hasegawa, ACS Nano {\bf 10}, 2761 (2016)
\bibitem{exp2SuperC} K. Li, X. Feng, W. Zhang, Y. Ou, L. Chen, Ke. He, Li-Li Wang, L. Guo, G. Liu, Qi-Kun Xue and X. Ma, Appl. Phys. Lett. {\bf 103}, 062601 (2013)
\bibitem{theory} M. Khantha, N. A. Cordero, L. M. Molina, J. A. Alonso, and L. A. Girifalco, Phys. Rev. B {\bf 70}, 125422 (2004)
\bibitem{lazic} P. Pervan, P. Lazi\'{c}, M. Petrovi\'{c}, I. \v{S}rut Raki\'{c}, I. Pletikosi\'{c}, M. Kralj, M. Milun and T. Valla, Phys. Rev. B {\bf 92}, 245415 (2015)
\bibitem{Nanolett-Dino} D. Novko, Nano Lett. {\bf 17}, 6991 (2017)
\bibitem{2Dplasmons} L. Maru\v{s}i\'{c} and V. Despoja, Phys.Rev B {\bf 95}, 201408(R) (2017)
\bibitem{IR3} F. Bonaccorso, Z. Sun, T. Hasan and A. C. Ferrari, Nature Photonics {\bf 4}, 611 (2010)
\bibitem{APP1} A. Vakil and N. Engheta, Science {\bf 332}, 1291 (2011) 
\bibitem{chinos}Wu Hua-Qiang, Linghu Chang-Yang, L ̈Hong-Ming, and Qian He, Chin. Phys. B {\bf 22}, 098106 (2013)
\bibitem{appl1} A. Pospischil, M. M. Furchi, and T. Mueller, Nat. Nanotechnol. {\bf 9}, 257 (2014)
\bibitem{appl3} J. S. Ross, P. Klement, A. M. Jones, N. J. Ghimire, J. Yan, D. G. Mandrus, T. Taniguchi, K. Watanabe, K. Kitamura, W. Yao, D. H. Cobden, and X. Xu, Nat. Nanotechnol. {\bf 9}, 268 (2014)
\bibitem{appl4} F. H. L. Koppens, T. Mueller, P. Avouris, A. C. Ferrari, M. S. Vitiello, and M. Polini, Nat. Nanotechnol. {\bf 9}, 780 (2014)
\bibitem{appl5} S. Jo, N. Ubrig, H. Berger, A. B. Kuzmenko, and A. F. Morpurgo, Nano Lett. {\bf 14}, 2019 (2014)
\bibitem{appl6} O. Lopez-Sanchez, E. Alarcon Llado, V. Koman, A. Fontcuberta i Morral, A. Radenovic, and A. Kis, ACS Nano {\bf 8}, 3042 (2014)
\bibitem{appl7} C.-H. Lee, G.-H. Lee, A. M. van der Zande, W. Chen, Y. Li, M. Han, X. Cui, G. Arefe, C. Nuckolls, T. F. Heinz, J. Guo, J. Hone, and P. Kim, Nat. Nanotechnol. {\bf 9}, 676 (2014)
\bibitem{appl8} B. W. H. Baugher, H. O. H. Churchill, Y. Yang, and P. Jarillo-Herrero, Nat. Nanotechnol. {\bf 9}, 262 (2014)
\bibitem{APP2} T. Low and P.  Avouris, ACS Nano {\bf 8}, 1086 (2014)
\bibitem{photopto}F. Bonaccorso, Z. Sun, T. Hasan, A. C. Ferrari, Nature Photonics {\bf 4}, 611 (2010)
\bibitem{appl2} L. Britnell, R. M. Ribeiro, A. Eckmann, R. Jalil, B. D. Belle, A. Mishchenko, Y. J. Kim, R. V. Gorbachev, T. Georgiou, S. V. Morozov, A. N. Grigorenko, A. K. Geim, C. Casiraghi, A. H. C. Neto, and K. S. Novoselov, Science {\bf 340}, 1311 (2013)
\bibitem{appl9}C. Zhu, D. Du, Y. Lin, 2D Mater. {\bf 2}, 032004 (2015), A. Y. Zhu, E. Cubukcu, 2D Mater. {\bf 2}, 032005 (2015)
\bibitem{stauber} T. Stauber and H. Kohler, Nano Lett. {\bf 16}, 6844 (2016)
\bibitem{Rukelj} V. Despoja, Z. Rukelj and L. Maru\v{s}i\'{c}, Phys.Rev B {\bf 94}, 165446 (2016)
\bibitem{Duncan2} V.Despoja, D. J. Mowbray, D. Vlahovi\' c and L. Maru\v si\' c, Phys. Rev. B {\bf 86}, 195429 (2012) 
\bibitem{wake} V. Despoja, K. Dekani\' c, M. \v Sunji\' c, and L. Maru\v si\' c, Phys. Rev. B {\bf 86}, 165419 (2012)
\bibitem{REELS}  Z.L.Wang, J.M.Cowley, Surface Science, {\bf 193}, 501 (1988)
\bibitem{Leo} L. Maru\v{s}i\'{c} and M. \v{S}unji\'{c}, Phys. Scr. {\bf 63}, 336 (2001)
\bibitem{QE} P. Giannozzi, S. Baroni, N. Bonini, M. Calandra, R. Car, C. Cavazzoni, D. Ceresoli, G. L. Chiarotti, M. Cococcioni, I. Dabo, {\em et.al.}, J. Phys.: Conden. Matter {\bf 21}, 395502 (2009)
\bibitem{pseudopotentials} N. Troullier and J. L. Martins, Phys. Rev. B {\bf 43}, 1993 (1991)
\bibitem{LDA} J.P. Perdew and A. Zunger,  Phys. Rev. B {\bf 23}, 5048 (1981)
\bibitem{lattice} R. Saito, G. Dresselhaus, and M. S. Dresselhaus, \textit{Physical Properties of Carbon Nanotubes} (Imperial College Press, London, (1998)
\bibitem{MPmesh} H.J. Monkhorst and J.D. Pack, Phys. Rev. B {\bf 13}, 5188 (1976)
\bibitem{DSV} D. Novko, M. \v Sunji\' c and V. Despoja, Phys. Rev B {\bf 93}, 125413 (2016)
\bibitem{MoS2} Z. Rukelj, A. \v Strkalj, V. Despoja, Phys. Rev. B {\bf 94}, 115428 (2016)
\end{thebibliography}
\end{document}